\documentclass[11pt]{article}
\usepackage[utf8]{inputenc}
\pdfoutput=1
\usepackage{amssymb,amsmath,mathrsfs,enumerate}
\usepackage{graphicx,rotate,multicol}
\usepackage{float}
\usepackage{tocloft}
\usepackage{subfig}
\usepackage[margin=10pt,labelfont=bf]{caption}
\usepackage{cite}
\usepackage{soul}
\usepackage[colorlinks=true,
            linkcolor=cyan,
            urlcolor=blue,
            citecolor=teal]{hyperref}
\usepackage{multirow}
\usepackage{physics}
\usepackage{xspace}

\makeatletter
\let\Hy@linktoc\Hy@linktoc@page
\makeatother

\usepackage{color}
\definecolor{ourcolor}{rgb}{0.7, 0.25, 0.05}
\usepackage{tikz,braket}
\long\def\rpl#1!!#2!!{\textcolor{red}{#1} \textcolor{blue}{#2}}

\def \order(#1){{\mathcal O} \left(#1 \right)}

\evensidemargin=\oddsidemargin
\allowdisplaybreaks

\usepackage{tikz,xcolor,hyperref}

\definecolor{lime}{HTML}{A6CE39}
\DeclareRobustCommand{\orcidicon}{\hspace{-1mm}
	\begin{tikzpicture}
	\draw[lime, fill=lime] (0,0) 
	circle [radius=0.16] 
	node[white] {{\fontfamily{qag}\selectfont \tiny \,ID}};
	\draw[white, fill=white] (-0.0525,0.095) 
	circle [radius=0.007];
	\end{tikzpicture}
	\hspace{-3mm}
}

\foreach \x in {A, ..., Z}{\expandafter\xdef\csname orcid\x\endcsname{\noexpand\href{https://orcid.org/\csname orcidauthor\x\endcsname}
			{\noexpand\orcidicon}}
}

\usepackage{xcolor}

\newcommand{\cevns}{CE$\nu$NS\xspace}

\title{\color{black}{\bf Neutrinos as background and signal in searches using the Migdal effect}}

\author {
\hspace{4pt}  Tarak Nath Maity$^{a,}$\footnote{tarak.maity.physics@gmail.com} \orcidA{} \\
 $^a$School of Physics, The University of Sydney, \\ ARC Centre of Excellence for Dark Matter Particle Physics, NSW 2006, Australia}

\date{}

\begin{document}

\maketitle

\begin{abstract}
Ionization or excitation resulting from the noninstantaneous response of the electron cloud to nuclear recoil is known as the Migdal effect. Dark matter searches utilizing this process set the most stringent bounds on the spin-independent dark matter-nucleon scattering cross section over a large region of the sub-GeV dark matter parameter space, underscoring its significance in dark matter detection. In this paper, we quantify the regions of dark matter parameter space that are challenging to probe via the Migdal effect due to the presence of dominant solar neutrino backgrounds for both liquid noble and semiconductor targets. Our findings reveal that there is no hard floor in the dark matter parameter space. Instead, we map the so-called neutrino fog. In mapping the neutrino fog, we identify the importance of incorporating the Migdal effect induced by neutrinos, as well as neutrino-electron scattering and dominant coherent neutrino-nucleus scattering, particularly for semiconductor targets. Furthermore, we demonstrate that a large portion of the relic density allowed parameter space lies within the neutrino fog. Finally, we estimate the exposure required to detect neutrino-induced Migdal events in direct detection experiments.             
\end{abstract}


\newpage

\hrule \hrule
\tableofcontents
\vskip 10pt
\hrule \hrule 

\section{Introduction}
\label{sec:intro}
A multitude of observations, spanning length scales from galactic and extragalactic to cosmological, indicate that approximately 85\% of the matter content of the Universe is composed of dark matter (DM) \cite{Cirelli:2024ssz}. While all these observational evidences of DM are gravitational in nature, the true nature of DM remains unknown. Among the various possibilities for uncovering the nature of DM, one approach is to search for its possible interactions with the visible Standard Model (SM) sector. These interactions between DM and SM particles can be investigated through a variety of methodologies, such as direct detection (DD) \cite{Lin:2019uvt, Cooley:2021rws}, indirect detection \cite{Slatyer:2021qgc}, and collider searches \cite{Buchmueller:2017qhf}. In particular, DD experiments measure the direct recoil of SM particles resulting from potential scattering between DM and SM states in clean underground laboratories.

In the last decade, DD experiments have made extraordinary progress, especially in probing weak-scale spin-independent (SI) DM-nucleon scattering \cite{XENON:2023cxc, PandaX:2024qfu, LZCollaboration:2024lux}. Searching for sub-GeV DM in DD calls for creative techniques, as energy deposition by such light DM often falls below the typical energy thresholds. These approaches include boosting nonrelativistic DM through its scattering with high energetic particles \cite{An:2017ojc, Bringmann:2018cvk, Ema:2018bih, Das:2021lcr, Ghosh:2024dqw}, the Migdal effect \cite{migdal1941ionization, Vergados:2005dpd, Ibe:2017yqa, Dolan:2017xbu, Bell:2019egg, Essig:2019xkx, Kang:2024kec, Bell:2023uvf}, scattering with kinematically favorable targets such as electrons instead of nuclei \cite{Essig:2011nj}, the use of novel condensed matter targets \cite{Kahn:2021ttr}, etc. Among these, the Migdal effect appears to be the current best probe of SI DM-nucleon scattering in the DM mass range of $\sim 0.01$--$1$\,GeV \cite{XENON:2019zpr, DarkSide:2022dhx, PandaX:2023xgl, SENSEI:2023zdf}. In the rest of this paper, we will focus exclusively on SI scattering and restrict our analysis to liquid noble targets, specifically liquid xenon (Xe) and liquid argon (Ar), and semiconductor targets, namely silicon (Si) and germanium (Ge).

The production of ionized electrons due to the nonadiabatic response of the electron cloud from a nuclear recoil is known as the Migdal effect\,\footnote{It should be noted that the Migdal effect has been experimentally observed for nuclear $\alpha$ decay \cite{Rapaport:1975zz, Rapaport:1975zza, PhysRevLett.34.173, PhysRevA.15.162} and $\beta$ decay \cite{PhysRev.93.518, PhysRevLett.108.243201}, but not from  scattering by neutral particles, such as in xenon by neutrons \cite{Xu:2023wev}. Experiments like MIGDAL \cite{MIGDAL:2022yip} aim to investigate this phenomenon.} (see Refs.\,\cite{GrillidiCortona:2020owp, Dey:2020sai, Nakamura:2020kex, Bell:2021zkr, Liao:2021yog, Bell:2021ihi, Liang:2022xbu, MIGDAL:2022yip, Blanco:2022pkt, Tomar:2022ofh,Chatterjee:2022gbo, Adams:2022zvg, Li:2022acp, Xu:2023wev, AtzoriCorona:2023ais, Gu:2023pfg, Li:2023xkf, Kahn:2024nyv} for related phenomenology).   While the nuclear recoil produced by ambient sub-GeV DM–nucleus scattering is small, ionization signatures from Migdal events can still be measured for relevant cross sections in DD experiments. Experiments using Xe and Ar targets, such as XENON1T \cite{XENON:2019zpr} and DarkSide-50 \cite{DarkSide:2022dhx} respectively, have set the most stringent limits on the SI DM-nucleon scattering cross section via the Migdal effect, employing ionization signal-only (S2-only) analyses for DM masses in the range of $\sim 0.03$--$1\,\mathrm{GeV}$. The SENSEI experiment, which uses Si target, has set the best limits in the DM mass range of $\sim 0.01$--$0.03\,\mathrm{GeV}$ by detecting electron/hole  produced via the Migdal effect \cite{SENSEI:2023zdf}. Throughout this paper, we broadly classify both types of searches—using liquid noble and semiconductor targets—as ionization-only analyses.  Interestingly, the large ton-scale Xe-based DD experiments have begun observing $^8$B solar neutrinos with a significance of approximately $3\sigma$ \cite{PandaX:2024muv, XENON:2024ijk}. Notably, this has been achieved even with the S2-only analysis \cite{PandaX:2024muv}. Motivated by this, and given that the Migdal effect provides the strongest bounds for DM masses in the range of $\sim 0.01$--$1$\,GeV, this paper explores regions of the SI DM-nucleon scattering parameter space where neutrinos, primarily solar neutrinos, would constitute significant backgrounds for DM searches using the Migdal effect. While this has previously been explored only for the Xe target in Ref.\,\cite{Herrera:2023xun}, in this work we extend the analysis in several directions: (a) we go beyond Xe by including additional target materials Ar, Si and Ge; (b) we employ a state-of-the-art analysis to more accurately predict the neutrino background-dominated DM parameter space—commonly referred to as the neutrino fog; (c) in addition to coherent elastic neutrino-nucleus scattering (\cevns) we also account for subdominant but relevant contributions from $\nu–e$ scattering and Migdal-induced events from \cevns, both of which play a crucial role in mapping the extent of the neutrino fog. Furthermore, we have also investigated the potential for observing the Migdal effect from neutrino interactions in future DD experiments.

As mentioned, in observing Migdal events from DM, the aforementioned experiments primarily measure ionization signals. While there could be many other sources for these events \cite{XENON:2019zpr, DAMIC:2021crr, SENSEI:2021hcn, Du:2023soy, Du:2020ldo}, neutrinos will always remain the ultimate irreducible background. Neutrinos can produce ionization through \cevns \cite{Freedman:1973yd} and neutrino-electron ($\nu-e$) scattering \cite{Marciano:2003eq}. Additionally, ionized electrons can also be generated from \cevns via the Migdal effect \cite{Ibe:2017yqa}. At large recoil energies, it is possible to distinguish between \cevns and $\nu-e$ scattering based on electron and nuclear recoils \cite{Aprile:2006kx, McGuire:2023cbg}. However, the topology of the Migdal events is challenging to categorize as either nuclear or electron recoil \cite{Dolan:2017xbu}. Moreover, since the current best limits are obtained using ionization-only analyses, distinguishing between these three types of events—\cevns, $ \nu-e $ scattering, and the Migdal effect—is difficult. Therefore, in our analysis, the total neutrino event rates are calculated by summing ionization contributions from \cevns, $ \nu-e $ scattering, and the Migdal effect from neutrinos. Using profile likelihood ratio techniques, we then map out regions of the DM parameter space that would be challenging to probe under this neutrino background. Following the formalism of Ref. \cite{OHare:2021utq}, we find that while neutrinos can slow down DM searches via the Migdal effect, they do not impose a hard boundary on the parameter space. Consequently, we map the so-called neutrino fog for the targets under consideration. We find that, for semiconductors, DM and neutrino-induced Migdal events share identical spectra within our DM mass range of interest, which would have led to a hard floor due to $pp$ solar neutrinos. However, the presence of indistinguishable $\nu-e$ events modifies the spectra, transforming the floor into a fog. This highlights the importance of including the subdominant $\nu-e$ background in mapping the neutrino fog. Additionally, we investigate the prospects of detecting Migdal events induced by neutrinos in future DD experiments.

The rest of the paper is organized as follows. In Sec.\,\ref{sec:DM-nu-events}, we calculate the DM and neutrino event rates. In Sec.\,\ref{sec:nufog}, we map the neutrino fog. Finally, in Sec.\,\ref{sec:nusignal}, we elucidate the exposure required to discover neutrino-induced Migdal scattering in DD, before concluding in Sec.\,\ref{sec:conclusion}.

\section{DM and neutrino event rates}
\label{sec:DM-nu-events}
In this section, we outline the relevant DM and neutrino event rates in DD experiments. In typical nuclear recoil searches, it is usually assumed that the electron cloud of the target immediately follows the recoiling nucleus. The recoiling nucleus then interacts with neighboring atoms, producing excitation and ionization, leading to observable signatures. However, scenarios for which electrons do not follow the nuclear recoil immediately, the hard scattering between DM-nucleus itself can produce ionizations  and excitations due to this quantum mechanical mismatch. This process is known as the Migdal effect.

For the liquid noble target, the event rates are calculated in the Migdal approximation. In this approach, the state of the electron cloud before the recoil is Galilean boosted in the rest frame of the recoiling atom. The boosted electron states are then combined with the ionized or excited electron cloud to obtain the ionization or excitation probability. While this effect has been known for decades \cite{migdal1941ionization}, a consistent calculation of the ionization probability from the Migdal effect was recently performed by Ref. \cite{Ibe:2017yqa}. For liquid noble targets, namely Xe and Ar, we utilize the electron ionization probability for isolated atoms given in Ref. \cite{Ibe:2017yqa} in our numerical calculations. For the DM mass range of interest, the velocity of the recoiling nucleus is small, and the dominant dipole contribution from the boost operator, as given in Ref.\,\cite{Ibe:2017yqa}, provides a good estimate of the event rate. As demonstrated in Ref.\,\cite{Cox:2022ekg}, higher-order terms contribute negligibly in ambient DM searches.

The calculation of the Migdal effect for semiconductor targets is more intricate \cite{Liang:2019nnx, Liu:2020pat, Knapen:2020aky, Liang:2020ryg} and cannot be performed in the same manner as for isolated atoms. This is due to the periodic structure of semiconductor crystals. The method of boosting in the nuclear rest frame creates a preferred rest frame, which differs from the usual crystal lattice rest frame, where the periodicity and other electronic properties are maintained. Furthermore, in a crystal, electrons experience the potential of neighboring ions, which makes them delocalized, unlike in the case of isolated atoms. To address this issue, we use the results of Ref. \cite{Knapen:2020aky}, which employs Bloch wave functions to account for the delocalization of electrons. Additionally, it treats the initial state of the nucleus as the ground state of the harmonic potential of the crystal and the final state of the nucleus as a plane wave. The plane wave approximation for the final state nucleus is valid only if the timescale of the DM-nuclear interaction (given by the inverse of the nuclear recoil energy) is shorter than the typical duration of lattice oscillations (determined by the phonon frequency, i.e., the inverse of the phonon energy). For DM masses roughly below $30\,\mathrm{MeV}$, the recoil energy is small enough to violate this condition, restricting the applicability of this framework to DM masses above $\sim 30\,\mathrm{MeV}$ (but see Ref.\,\cite{Berghaus:2022pbu}). In our numerical calculations, we utilize the GPAW ionization probability from DarkELF \cite{Knapen:2021bwg} due to its reliability, even for single-electron thresholds.

Having outlined the generic formalism required to calculate the Migdal event rates, we now proceed to compute the DM event rates arising from the Migdal effect, as well as all the relevant neutrino event rates.

\subsection{DM event rates}
\label{subsec:DM-events}
For DM mass $m_\chi$ and SI DM-nucleon cross section $\sigma_{\rm SI}$, the differential event rate from the Migdal effect is given by
\begin{equation}
\frac{\mathrm{d}^2 R_{\chi}}{\mathrm{d} E_N \, \mathrm{d} E_e} = \mathcal{N} \rho_{\chi}\frac{C \sigma_{\rm SI}}{2 m_{\chi} \mu^2_{\chi n}} F^2(E_N) \dv{P}{E_e} g_{\chi}(v_{\rm min}),
\label{eq:DMrate}
\end{equation}
where $\mathcal{N}$ is the exposure, $\rho_{\chi} = 0.3\, \rm{GeV/cm}^{3}$ is the local DM density, and $C$ ($=A^2$, with target atomic number $A$) represents the coherence enhancement factor for SI scattering. The DM-nucleon  reduced mass is $\mu_{\chi n}$, and $F(E_N)$ is the Helm form factor with nuclear recoil energy $E_N$. The differential Migdal ionization probability is $\mathrm{d}P/\mathrm{d}E_e$, with $E_e$ denoting the energy of the ionized electron. For Xe and Ar targets, we use the ionization probabilities from Ref.\,\cite{Ibe:2017yqa} along with the quoted normalizations, while for semiconductor targets, we adopt the same from Ref.\,\cite{Knapen:2021bwg}. The mean inverse speed, $g_{\chi}(v_{\rm min})$, is given by
\begin{equation}
g_{\chi}(v_{\rm min}) = \int_{v_{\rm min}} \frac{f^{\rm gal}_{\chi}(\mathbf{v+v}_{\rm E})}{v} d^3v,
\end{equation}
where Earth's velocity in the Galactic rest frame is $\mathbf{v}_{\rm E}$, which is taken as its average value, neglecting any time dependency. The DM speed distribution in the Galactic rest frame is $f^{\rm gal}_{\chi}(\mathbf{v})$, which we assume to be a Maxwell-Boltzmann distribution truncated at the escape velocity $v_{\rm esc} = 528\,$km/s, with the velocity dispersion related to the local circular velocity $v_0 = 233\,$km/s \cite{Evans:2018bqy, Maity:2022enp, Maity:2020wic}. The minimum DM velocity, $v_{\rm min}$, required to produce nuclear recoil energy $E_N$ and electron ionization energy $E_e$, is given by \cite{Dolan:2017xbu}
\begin{equation}
v_{\rm min} = \sqrt{\frac{m_N E_N}{2 \mu_{\chi N}^2}} + \frac{E_e + \Delta E}{\sqrt{2 m_N E_N}},
\end{equation} 
where $m_N$ is the nuclear mass and $\mu_{\chi N}$ is the DM-nucleus reduced mass. For liquid noble targets, $\Delta E$ is the binding energies of the shell under consideration, whereas for semiconductor targets $\Delta E=0$, since the electronic structure is described by energy bands rather than by atomic shells, as in the case of noble liquids. We will account for the band gap energies later. For the rest of the paper we will follow this notation.

In our numerical analysis, we aim to adopt an ionization-only analysis methodology by calculating the event rate against the number of electrons (a proxy for ionized energy) produced due to the Migdal effect. This approach is motivated by the fact that the best limits in our region of DM parameter space of interest come from ionization-only analyses. For example, the XENON1T S2-only analysis sets the strongest constraint on the DM mass range $\sim 0.6\,$--$1\,$GeV using the Migdal effect \cite{XENON:2019zpr}. A similar ionization-only analysis from DarkSide-50 sets the strongest bounds in the DM mass range $\sim 0.03\,$--$0.6\,$GeV \cite{DarkSide:2022dhx}. The current results from SENSEI set the strongest bound for the DM mass range below $\sim 0.03\,$GeV \cite{SENSEI:2023zdf}. All of these analyses are blind in separating nuclear and electron recoil events.

To obtain the event rates as a function of $n_e$, we calculate the detected energy, $E_{\rm det} = E_e + \Delta E + Q_f E_N$, where $Q_f$ is the quenching factor. We use the Lindhard quenching model \cite{lindhard1963integral} for Xe and Ar, whereas for Si and Ge, we use the fiducial model from Ref.\,\cite{Essig:2018tss}. Any other choice of quenching model would lead to some uncertainty in the DM rate; however, this is expected to be small, as the nuclear recoil energy is considerably smaller than the electron recoil energy. We integrate Eq.\,\eqref{eq:DMrate} over $E_N$ in the accessible phase space regions to obtain the differential rate as a function of $E_{\rm det}$. For Xe, we use the model from Ref.\,\cite{Baxter:2019pnz}, and for Ar, we use ionization model of Ref.\,\cite{DarkSide:2021bnz} with method from\,\cite{DarkSide-50:2023fcw} to convert $E_{\rm det}$ to $n_e$. For semiconductor targets, we use the model where one extra $n_e$ is produced for every additional mean energy deposition above the band gap energy\,\cite{Essig:2015cda}. The corresponding event rates are depicted in Figs.\,\ref{fig:rate_xe_ar} and \ref{fig:rate_si_ge} for liquid noble and semiconductor targets, respectively, with two representative choices of DM masses and cross sections. The rise in the event rate above a certain $n_e$ for liquid noble targets is an artifact of recoil energies exceeding the binding energies of a particular electron shell.

\subsection{Neutrino event rates}
\label{subsec:nu-events}
Next, we turn to the calculation of neutrino event rates. The relevant neutrino sources include the Sun, Earth, diffuse supernova neutrinos, and atmospheric neutrinos. In our numerical calculations, we consider all these sources. However, solar neutrinos dominate due to their high flux compared to other sources and their sufficient energy to produce relevant recoils. The neutrino fluxes and their uncertainties are adopted from Ref.\,\cite{OHare:2020lva}. These neutrinos produce ionization through three processes: \cevns, neutrino-electron scattering, and the  Migdal effect. We consider all these events as in the low-energy regime where most Migdal effect induced DM events occur, current experimental methodologies cannot distinguish whether ionized electrons originate from nuclear recoil, electron recoil, or Migdal-like events\footnote{At higher recoil energies, it becomes possible to separate nuclear and electron recoils. However, the percentage contribution of both the DM signal and the total neutrino background at such energies is very minimal.}. The mechanisms for ionization production differ between \cevns and the Migdal effect. In \cevns, neutrino interactions induce nuclear recoil, which subsequently interacts with neighboring atoms via inelastic collisions or with electrons through Coulomb forces, leading to ionization. This process is often described in the literature as ionization production from nuclear recoil via quenching. In contrast, the Migdal effect produces most of the ionized electrons directly from the hard inelastic scattering process itself.

\begin{figure*}[t]
\begin{center}
	\subfloat[\label{sf:rate_xe}]{\includegraphics[angle=0.0,width=0.45\textwidth]{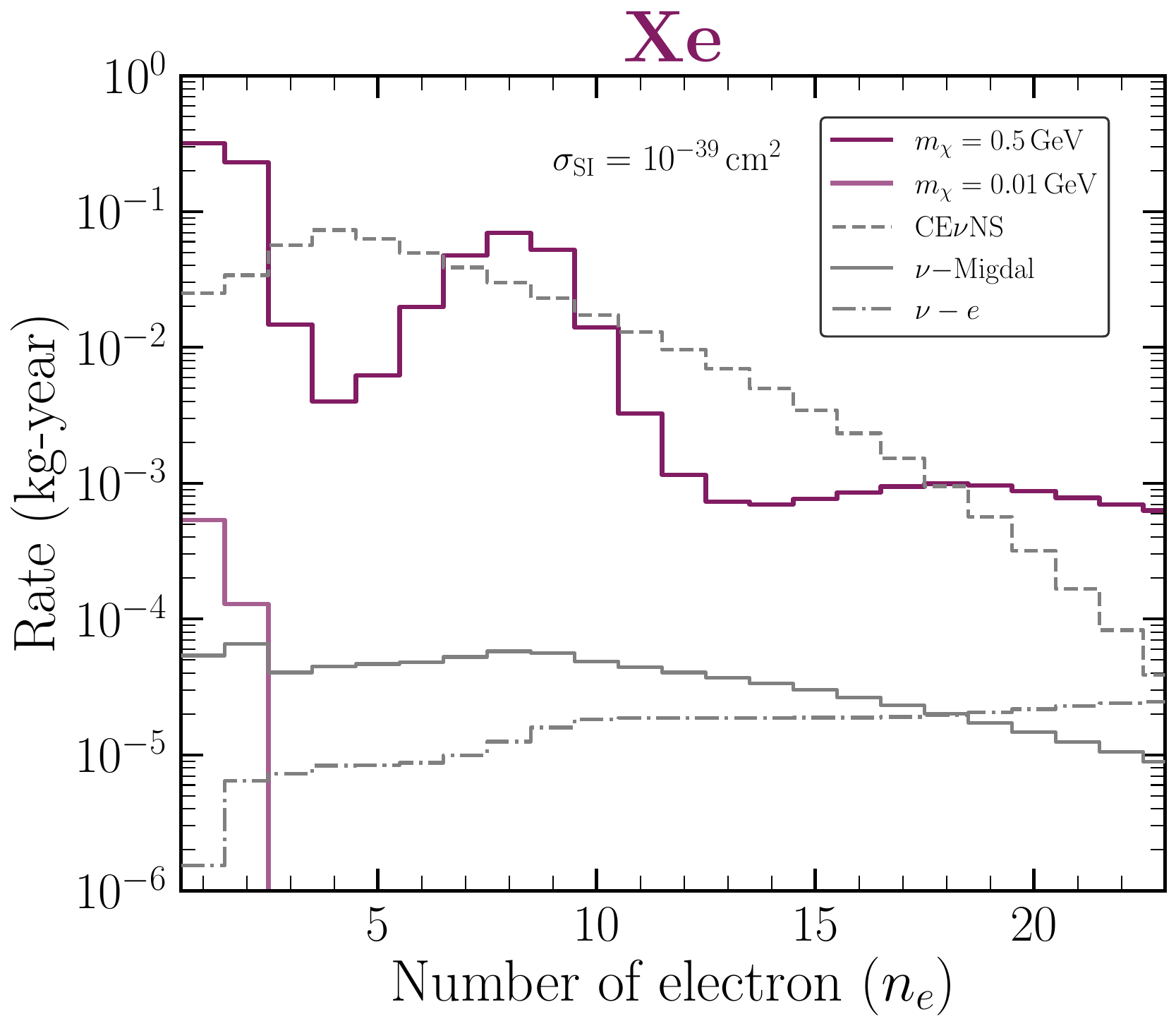}}~~
	\subfloat[\label{sf:rate_ar}]{\includegraphics[angle=0.0,width=0.45\textwidth]{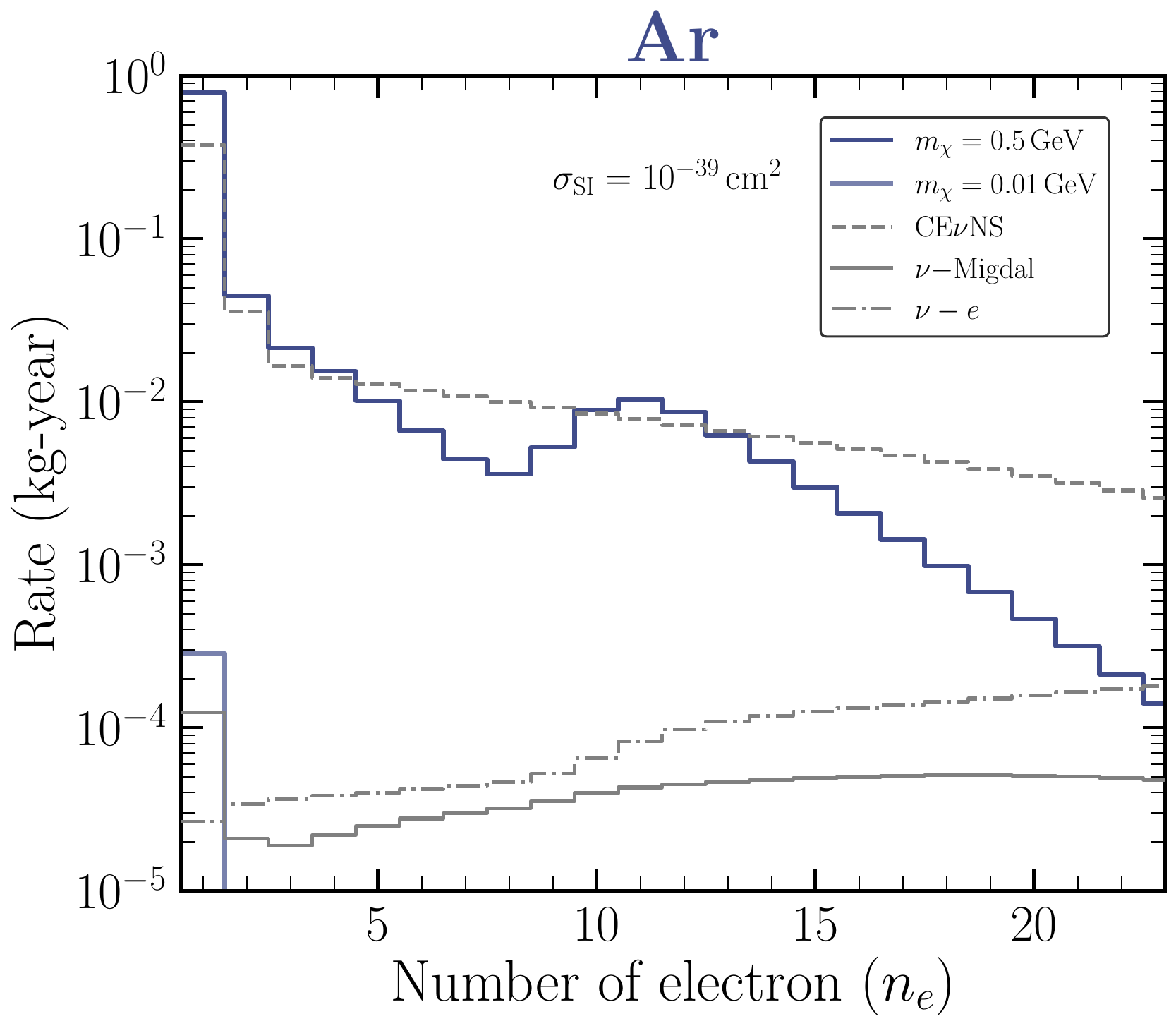}}~~\\	
	\caption{Event rate for DM and neutrinos as a function of the number of electrons for $1\,\text{kg-year}$ exposure. In both panels, the \cevns, $\nu-e$, and neutrino-induced Migdal events are represented by the gray dashed, dot-dashed, and solid lines, respectively. For both panels, the SI DM-nucleon cross section is assumed to be $10^{-39}\,{\rm cm}^2$. (a) For the Xe target: The purple and light purple solid lines correspond to DM masses of $0.5$ and $0.01\,\text{GeV}$, respectively. (b) For the Ar target: The blue and light blue solid lines correspond to DM masses of $0.5$ and $0.01\,\text{GeV}$, respectively.}
	\label{fig:rate_xe_ar}
\end{center}	
\end{figure*}

The differential \cevns or $\nu-e$ induced event rates can be expressed as
\begin{equation}
\dv{R_{\nu j}}{E_j} =  \mathcal{N} \int_{E_{\nu, j}^{\,\text{min}}}\dv{\sigma_{\nu j}}{E_j}\dv{\phi_\nu}{E_\nu}\mathrm{d}E_\nu.
\label{eq:cenuns-nue-rate}
\end{equation}
\begin{figure*}[t]
\begin{center}
	\subfloat[\label{sf:rate_si}]{\includegraphics[angle=0.0,width=0.45\textwidth]{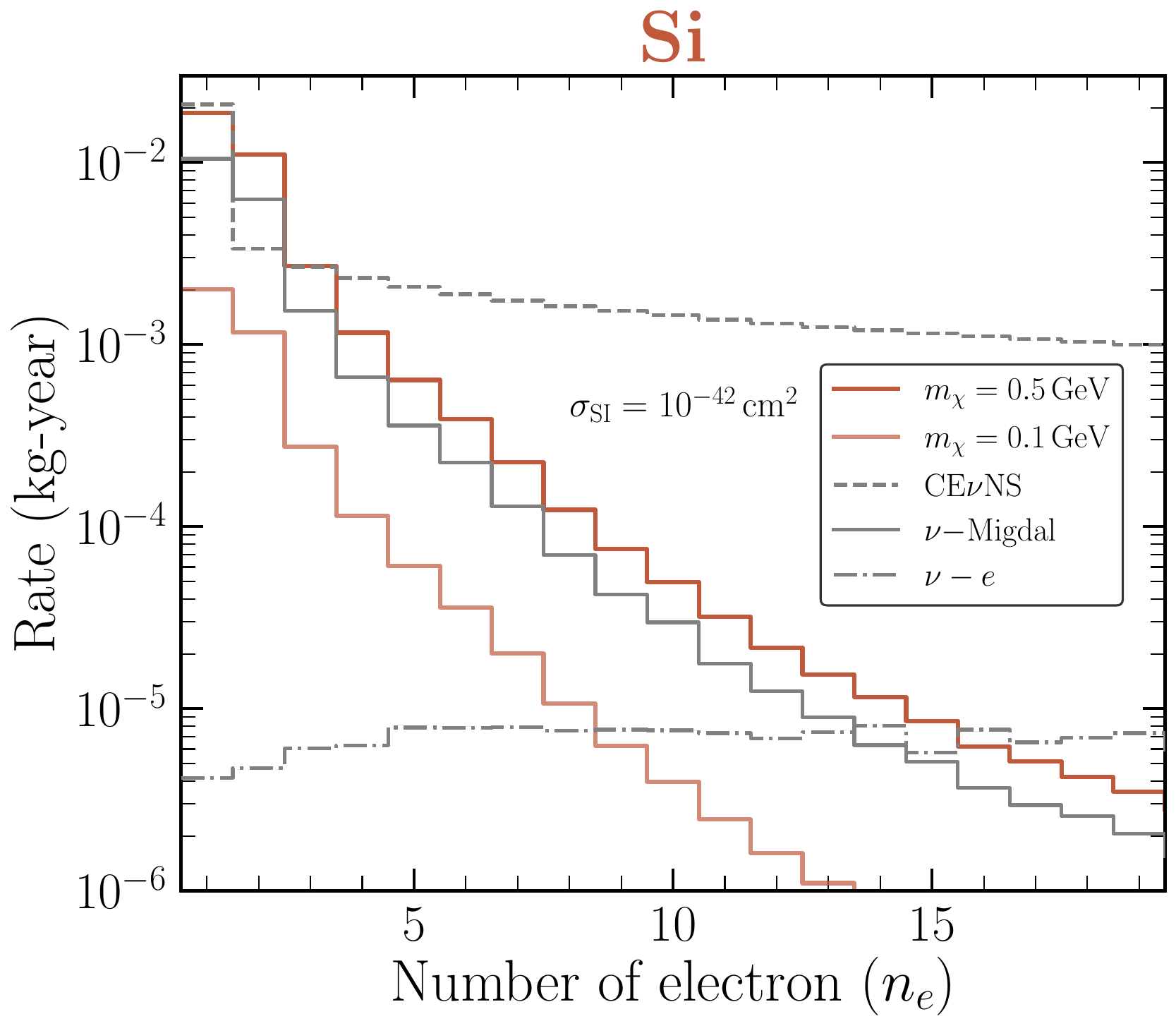}}~~
	\subfloat[\label{sf:rate_ge}]{\includegraphics[angle=0.0,width=0.45\textwidth]{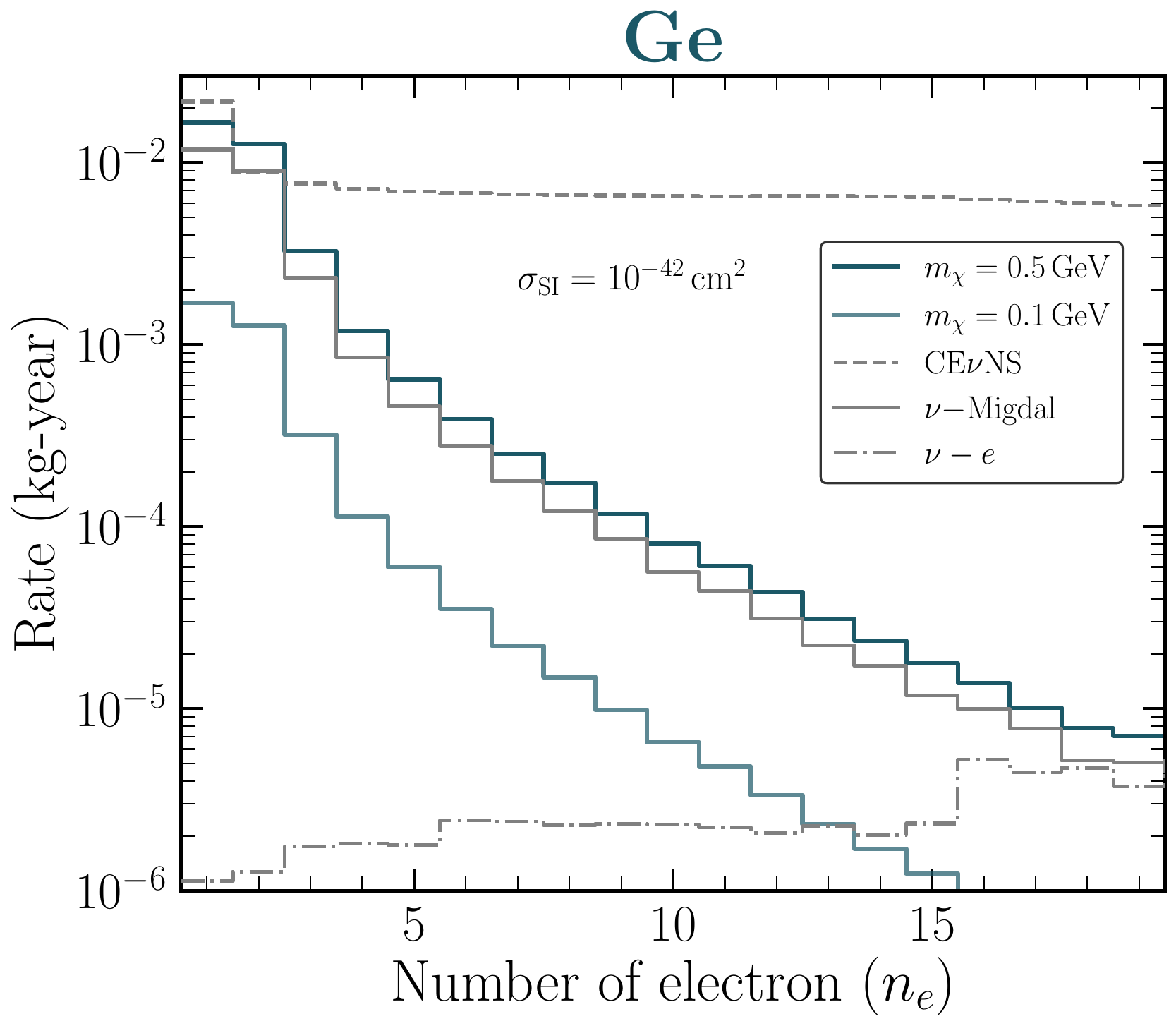}}~~\\	
	\caption{Same as Fig.\,\ref{fig:rate_xe_ar}, but for Si and Ge targets in (a) and (b), respectively. The chosen DM masses are $0.5$ and $0.1\,\text{GeV}$ with SI DM-nucleon cross section of $10^{-42}\,{\rm cm}^2$.} 
	\label{fig:rate_si_ge}
\end{center}	
\end{figure*}
The index $j$ corresponds to $N$ or $e$ for \cevns and $\nu-e$, respectively. The neutrino flux is ${\rm d}\phi_\nu/{\rm d}E_\nu$, with $E_\nu$ representing the neutrino energy. The minimum neutrino energy required to produce a recoil of $E_j$ is denoted by $E_{\nu,j}^{\,\rm min}$. The differential \cevns cross section can be read as
%
\begin{equation}
\dv{\sigma_{\nu N}}{E_{N}} = \frac{G_F^2}{4\pi}Q_W^2 m_N \left(1-\frac{m_N E_N}{2E_\nu^2}\right)F^2(E_N),
\label{eq:cenuns-x-section}
\end{equation}
with the Fermi constant $G_F$, the weak nuclear hypercharge for a target with $\mathbb{N}$ neutrons and $\mathbb{Z}$ protons is given by $Q_W = \mathbb{N} - \mathbb{Z}(1 - 4\sin^2\theta_W)$. The weak mixing angle $\theta_W$ is fixed to its low-energy value as given in Ref.\,\cite{Erler:2017knj}. By substituting Eq.\,\eqref{eq:cenuns-x-section} into Eq.\,\eqref{eq:cenuns-nue-rate}, along with the relevant neutrino flux, the differential event rate as a function of $E_N$ can be obtained. This is subsequently converted into event numbers as a function of electrons, following the prescription outlined in Ref.\,\cite{Essig:2018tss}. The corresponding \cevns induced ionization event rates for liquid noble and semiconductor targets are represented by the gray dashed lines in Figs.\,\ref{fig:rate_xe_ar} and \ref{fig:rate_si_ge}, respectively.

For neutrino-electron scattering event rates, we use the differential cross section mentioned below, along with replacing $E_j$ by $E_e$ in Eq.\,\eqref{eq:cenuns-nue-rate},
\begin{equation}
\dv{\sigma_{\nu e}}{E_e} = P_{\nu_e \nu_e} \dv{\sigma_{\nu_e e}}{E_e} + \sum_{l = \mu, \tau} P_{\nu_e \nu_l} \dv{\sigma_{\nu_l e}}{E_e}.
\label{eq:nu_e_xsection}
\end{equation}
Here, the probability of electron neutrino flavor oscillation to itself and other flavors is denoted by $P$'s and adapted from \cite{ParticleDataGroup:2022pth} for normal ordering. The differential flavor-dependent neutrino-electron scattering cross section is given by
\begin{align}
\dv{\sigma_{\nu_l e}}{E_e} =  Z_{\rm eff}(E_e) \frac{G_F^2 m_e}{2\pi} \left[ \left(g_A^{\nu_l} +  g_V^{\nu_l}\right)^2 +  \left(g_A^{\nu_l} - g_V^{\nu_l}\right)^2  \right. 
 \left.  \left(1- \frac{E_e}{E_\nu} \right)^2 + \left({g_A^{\nu_l}}^2 - {g_V^{\nu_l}}^2\right) \frac{m_e E_e}{E^2_{\nu}} \right]\, ,
\end{align}
where $Z_{\rm eff}(E_e)$ is the recoil energy-dependent effective charge, adapted from the electron configurations provided in Ref.\,\cite{Cox:2022ekg}. Following Ref.\,\cite{Brown2006IntensityOD}, for semiconductors, we extend $Z_{\rm eff}$ up to the band gap energies. The axial and vector couplings are denoted by $g_A^{\nu_l}$ and $g_V^{\nu_l}$, respectively, with formulas provided in Ref.\,\cite{Maity:2024aji}. After deriving the differential recoil rate with respect to $E_e$, we convert it with respect to $n_e$ using the models from Ref.\,\cite{Baxter:2019pnz} for Xe, Refs.\,\cite{DarkSide-50:2023fcw, DarkSide:2021bnz} for Ar, and Ref.\,\cite{Essig:2015cda} for semiconductors. The $\nu-e$ event rate is represented by the gray dot-dashed lines in Figs.\,\ref{fig:rate_xe_ar} and \ref{fig:rate_si_ge} for liquid noble and semiconductor targets, respectively. 

Finally, we discuss the ionization events that arise due to the Migdal effect from \cevns. These are obtained using the following differential event rate:
\begin{equation}
\frac{\mathrm{d} R_{\nu}}{\mathrm{d} E_e} = \mathcal{N} \int \int \dv{\sigma_{\nu N}}{E_N} \dv{P}{E_e} \dv{\phi_\nu}{E_\nu} \mathrm{d}E_\nu \, \mathrm{d} E_N,
\label{eq:nu_migdal_rate}
\end{equation}
where ${\rm d}{\sigma_{\nu N}}/{\rm d}{E_N}$ can be obtained from Eq.\,\eqref{eq:cenuns-x-section}. The nuclear recoil energy $E_N$ and the neutrino energy $E_\nu$ are integrated over the kinematically allowed accessible region\,\cite{Bell:2019egg}
\begin{align}
 & \frac{(E_e + \Delta E)^2}{2 m_N} \leq E_N \leq \frac{\left( 2 E_\nu - (E_e + \Delta E)\right)^2}{2 (m_N + 2 E_\nu)}, \\
 & E_{\nu} \geq \frac{1}{2}\left(E_N + E_{e} + \Delta E + \sqrt{E_N^2 + 2 E_N M_N + 2 E_N (E_{e} + \Delta E)}\right),
\label{eq:Enr_limit_migdal}
\end{align}
where $\Delta E = E_{nl}$ is the binding energy of each shell for Xe and Ar, and $\Delta E = 0$ for Si and Ge. For liquid noble targets, there is an implicit summation over atomic shells since the differential ionization probabilities and $\Delta E$ depend on the atomic shells. Following Eq.\,\eqref{eq:nu_migdal_rate}, we obtain the differential rate with respect to the detected energy $E_{\rm det} = E_e + \Delta E + Q_f E_N$, which is then converted into the number of electrons using the procedure mentioned in the case of DM. The neutrino event rate from the Migdal effect is shown by the solid gray lines in Fig.\,\ref{fig:rate_xe_ar} for Xe and Ar and in Fig.\,\ref{fig:rate_si_ge} for Si and Ge.

Interestingly, neutrino-induced Migdal rates are larger for semiconductors compared to liquid noble targets and for the first few electron bins, they are comparable to the \cevns rate. The larger rate for semiconductors is due to the higher ionization probabilities and $\mathcal{O}$(eV) band gap energy. This can also be understood from Fig. 2 of Ref.\,\cite{Knapen:2021bwg}, where the authors compared the reach in DM parameter space for a $100$ kg-year Xe target (obtained from Ref.\,\cite{Essig:2019xkx}, whose ionization probabilities are roughly comparable to those used in this work) with $1$ kg-year of Si and Ge. For most of the DM mass range, the reach for Xe targets, even with $100$ times more exposure than for semiconductors, is nearly $10$ times weaker. This implies that the total rate for semiconductors would be approximately $10^3$ times higher than for Xe. Thus, the use of the ionization probability from Ref.\,\cite{Knapen:2021bwg} in the neutrino Migdal rate calculation makes the neutrino-induced Migdal event rate orders of magnitude larger than for liquid noble targets, pushing it to the level of quenched \cevns for the first few bins. The almost identical shapes of the event rate for displayed DM masses and neutrino Migdal rates for semiconductors in Fig.\,\ref{fig:rate_si_ge} are an artifact of using the same ionization probability and $\mathcal{O}$(eV) band gap which could be accessible by nonrelativistic DM and relativistic neutrinos. While we use the same ionization probability for calculating DM and neutrino event rates for liquid noble targets, lighter DM may not have enough energy to ionize an electron from inner shells, whereas the relevant neutrino could. This leads to a difference in the event rate shape.       
    
\section{Neutrino fog for Migdal search}
\label{sec:nufog}

With the DM and neutrino event rates given in the previous section, in this section we calculate the region of the SI DM-nucleon parameter space that would be difficult to probe via Migdal effect in DD experiments due to the presence of these neutrino backgrounds. For each $n_e$ bin, we add up the contributions from the three types of neutrino events, i.e., \cevns, $\nu-e$ scattering, and Migdal-induced neutrino events, to obtain the total rate in each bin, given that it would be difficult to differentiate between these categories with ionization-only analysis. Like in other neutrino background analyses \cite{Monroe:2007xp, Strigari:2009bq, Gutlein:2010tq, Billard:2013qya, Gutlein:2014gma, OHare:2016pjy, OHare:2020lva, Wyenberg:2018eyv, Herrera:2023xun, Carew:2023qrj}  in DM searches, we do not consider other experimental backgrounds here. This will lead to a conservative estimate, as the inclusion of other backgrounds would worsen the chances of detecting smaller cross sections.

In our statistical analysis we employ the profile likelihood ratio techniques and follow the procedure of Ref.\,\cite{OHare:2021utq} to obtain minimum SI DM-nucleon cross section that can be probed (called discovery limit) for a given exposure and DM mass. Numerically, we exploit the following binned likelihood for a generic model $\mathcal{M}^{\nu}_{\sigma_{\rm SI}}$:  
\begin{equation}
\mathcal{L}\left(m_{\chi},\sigma_{\rm SI},\Phi|\mathcal{M}^{\nu}_{\sigma_{\rm SI}}\right) = \prod_{i=1}^{N_e} \mathcal{P}\left(N^{\rm obs}_{i} | N^{\chi}_{i} + \sum_{j=1}^{n_\nu}\sum_{k=1}^{3} N^{\nu}_{ik}(\phi_j)\right) \prod_{j=1}^{n_{\nu}} \mathcal{G}(\phi_j),
\label{eq:ll}
\end{equation}  
where $\Phi = \{\phi_1, \dots, \phi_{n_{\nu}}\}$ represents the flux normalization for all the considered $n_\nu$ types of neutrino sources. The Poisson probability is denoted as $\mathcal{P}$. In the $i^{\rm th}$ electron number bin, the observed event is $N^{\rm obs}_{i}$, and the event from DM is $N^{\chi}_{i}$. For each neutrino source, there are three categories of events contributing to the total neutrino events in each bin; thus, we sum over those (with $k$ running from 1 to 3) to get the total number of neutrino events in the $i^{\rm th}$ bin. The total number of electron bins included in our analysis is denoted by $N_e$. The Gaussian function $\mathcal{G}$ accounts for the uncertainty in the neutrino flux and is given by
\begin{equation}
\mathcal{G}(\phi_j) = \frac{1}{\sqrt{2 \pi} \sigma_j}\exp{- \frac{1}{2}\left(\frac{\phi - \phi_j}{\sigma_j}\right)^2},
\label{eq:gaussian}
\end{equation}  
where the neutrino flux normalization for the $j^{\rm th}$ type is $\phi_j$, and the uncertainty in the flux is $\sigma_j$. The numerical values of the flux normalizations and uncertainties are adopted from \cite{OHare:2020lva}.  

With the likelihood function given in Eq.\,\eqref{eq:ll}, our aim is to construct the test statistic that allows us to statistically compare the background-only model ($\mathcal{M}^{\nu}_{\sigma_{\rm SI} = 0}$) with the signal and background model ($\mathcal{M}^{\nu}_{\sigma_{\rm SI}}$). The background-only model consists only of neutrinos, in other words, by setting $\sigma_{\rm SI} = 0$ (i.e., $N^{\chi}_{i}=0$) in Eq.\,\eqref{eq:ll}. The signal and background model includes both the signal DM and the background neutrino contributions. The corresponding likelihood ratio is then defined as
\begin{equation}
\Lambda = \frac{\mathcal{L}(\sigma_{\rm SI}=0,\Phi^{\prime \prime}|\mathcal{M}^{\nu}_{\sigma_{\rm SI}=0})}{\mathcal{L}(\sigma_{\rm SI},\Phi^{\prime}|\mathcal{M}^{\nu}_{\sigma_{\rm SI}})},
\label{eq:lambda}
\end{equation}
where we omitted quoting $m_{\chi}$ since the above likelihood ratio is calculated for a fixed $m_{\chi}$ and then iterated over the DM mass range of interest. In Eq.\,\eqref{eq:lambda}, we maximize the likelihood $\mathcal{L}$ over the uncertainties in the neutrino flux to find $\Phi^{\prime \prime}$ and $\Phi^{\prime}$, where $\mathcal{L}$ is maximized for the $\mathcal{M}^{\nu}_{\sigma_{\rm SI} = 0}$ and $\mathcal{M}^{\nu}_{\sigma_{\rm SI}}$ models, respectively. Finally, we construct the test statistic (TS) for positive values of $\sigma_{\rm SI}$ as follows:
\begin{equation}
q_0 = -2 \ln \Lambda
\label{eq:TS}
\end{equation}
\begin{figure*}[t]
\begin{center}
	\subfloat{\includegraphics[angle=0.0,width=0.55\textwidth]{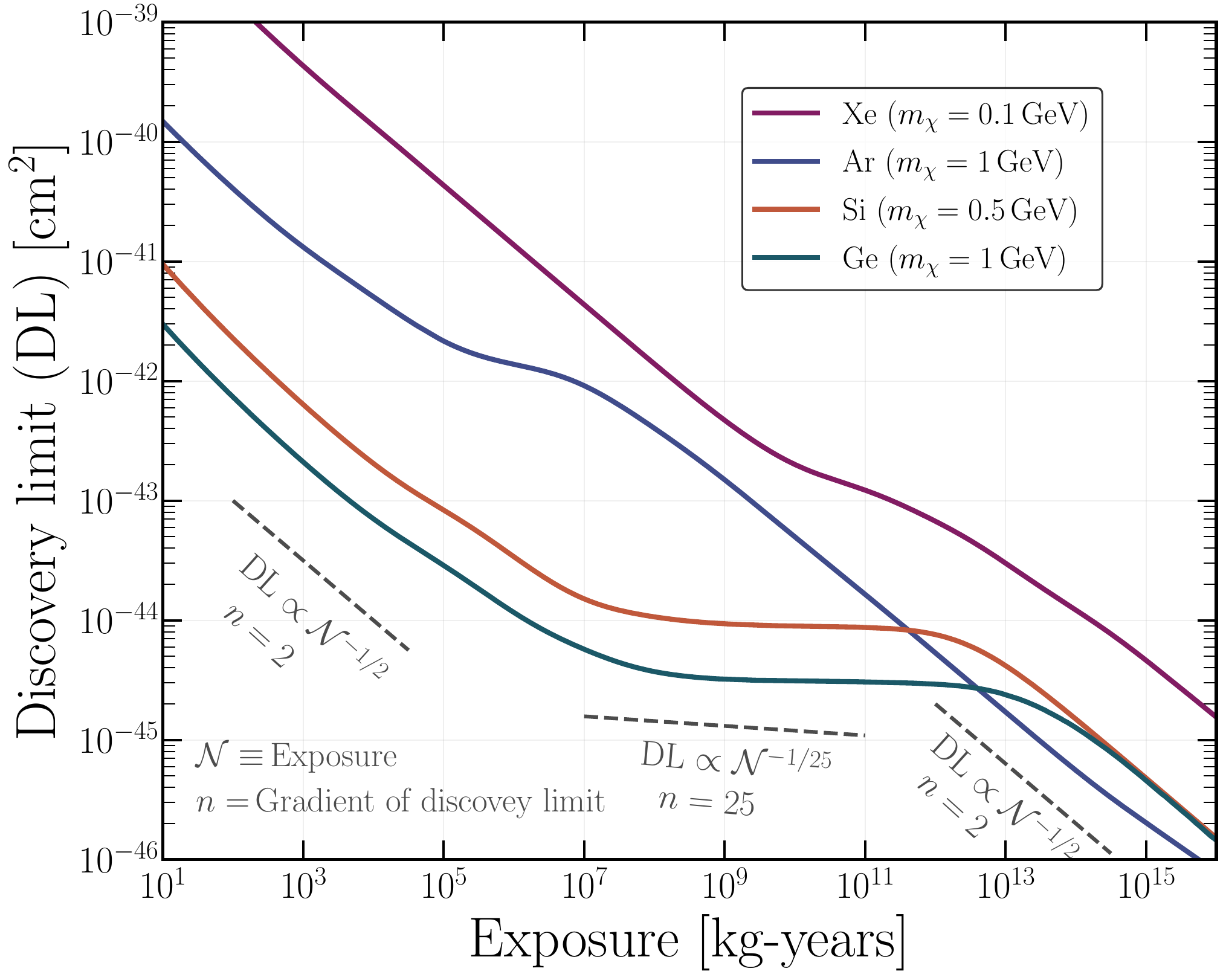}}	
	\caption{Evolution of the discovery limit with exposure. The solid purple,  blue, orange, and green lines represent Xe, Ar, Si, and Ge targets, respectively. The chosen DM masses for Xe, Ar, Si, and Ge are $0.1$, $1$, $0.5$, and $1\,\text{GeV}$, respectively. The indicative behavior of the DL with exposure is also represented by the gray dashed lines.}
	\label{fig:DL_1D_all}
\end{center}	
\end{figure*}
Since our signal and background models differ only by one parameter, $\sigma_{\rm SI}$, according to Wilks's theorem, the TS will asymptotically follow a $\chi^2_1$ distribution. The significance of testing our signal model against the background is given by $\sqrt{q_0}$. In our numerical evaluation, for a fixed DM mass and exposure, we iterate over $\sigma_{\rm SI}$ to find the cross section where $q_0$ becomes equal to $9$, which corresponds to a $3\sigma$ discovery limit. Note that, in doing this, we utilize the Asimov dataset, which assumes the observation exactly matches the expectation from a model, i.e., $N^{\rm obs}_{i} = N^{\rm exp}_{i}$ for all $i$ in Eq.\,\eqref{eq:ll}. Instead of conducting a dedicated Monte Carlo simulation, this is a standard practice in the literature to economize computational cost.

Using the statistical methodology discussed above, we calculate the evolution of the discovery limit (DL) with exposure. This is shown in Fig.\,\ref{fig:DL_1D_all} for all the considered targets. In Fig.\,\ref{fig:DL_1D_all}, the choices of DM masses vary to maintain clarity in the plot. For small exposures, the discovery limit roughly scales with the inverse of the exposure, as neutrinos do not significantly contribute due to the limited exposure. As the exposure increases, neutrinos begin to appear, and the discovery limit enters the regime dominated by statistical uncertainties, where it scales as $\propto 1/\sqrt{\mathcal{N}}$, with $\mathcal{N}$ being the exposure.
\begin{figure*}[t]
\begin{center}
	\subfloat[\label{sf:DL_2D_Xe}]{\includegraphics[angle=0.0,width=0.45\textwidth]{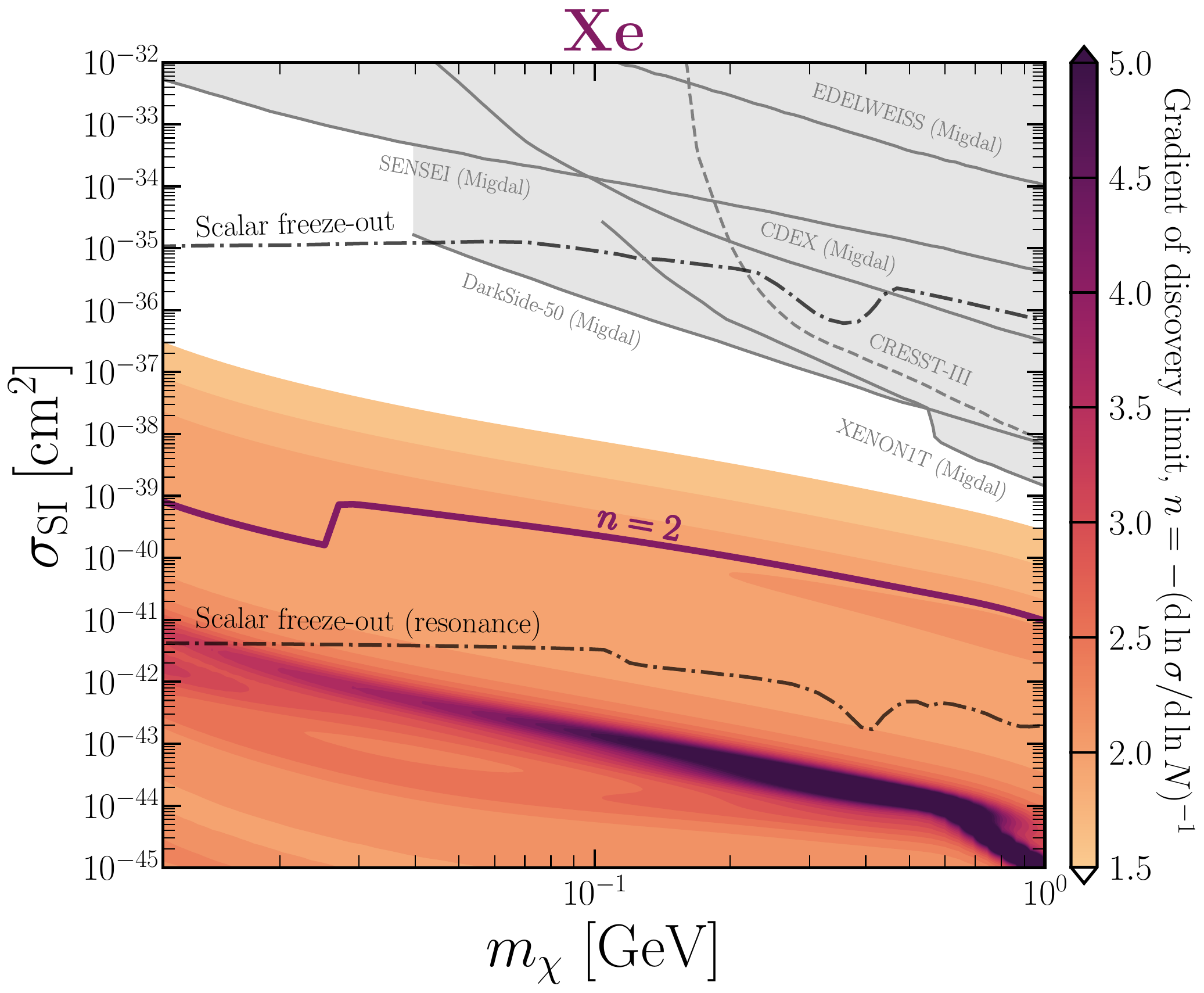}}~~
	\subfloat[\label{sf:DL_2D_Ar}]{\includegraphics[angle=0.0,width=0.45\textwidth]{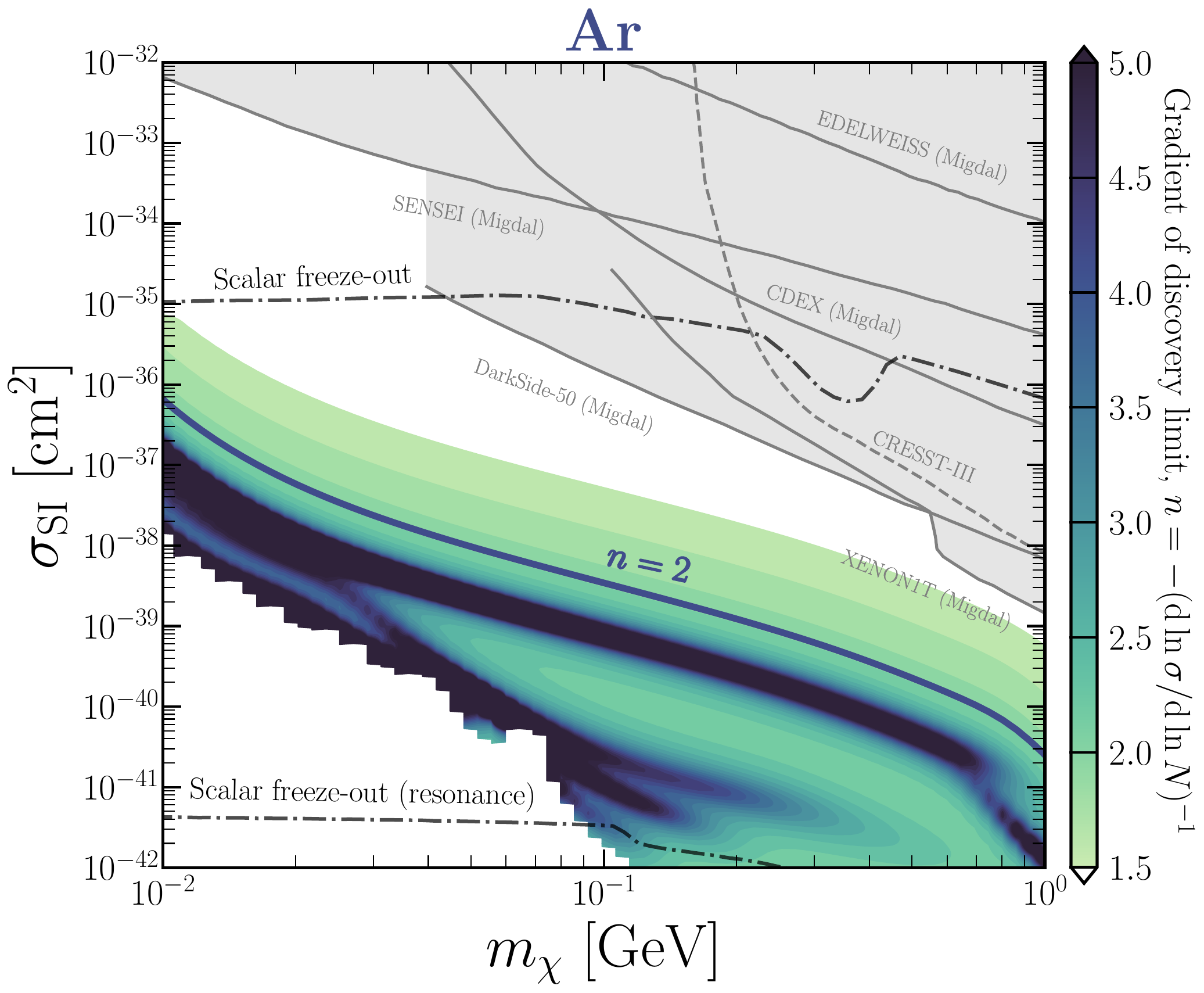}}~~\\	
	\caption{Neutrino fog map with $n$ as the gradient of the DL in SI DM-nucleon and DM mass parameter space. The region between black dot-dashed lines corresponds to scalar DM satisfying relic density. The gray-shaded regions are ruled out by current observations. The transition boundary between neutrino statistical and systematic uncertainty-dominated regimes is denoted by $n=2$. (a) For the Xe target, whereas (b) is for the Ar target. For the Ar target, the required exposure for the blank region below the color map exceeds $10^{15}\,$kg-years.}
	\label{fig:DL_2D_Xe_Ar}
\end{center}	
\end{figure*}
Further increasing the exposure causes the discovery limit to enter the regime dominated by systematic uncertainties. Here, the DM signal produces fewer events than the fluctuations in the background due to uncertainties in the knowledge of the neutrino background. In this regime, the discovery limit scales as $\propto \sqrt{(1+\mathcal{N}\delta\Phi^2)/\mathcal{N}}$, where $\delta \Phi$ is the uncertainties in the neutrino fluxes. This behavior is illustrated by the flat region and the region adjacent to it in each curve of Fig.\,\ref{fig:DL_1D_all}.

Beyond this regime, increasing exposure further enables experiments to measure their own background (neutrinos, in this case) and use energy spectral information to distinguish between DM and neutrinos. This differentiation is only possible if the DM and neutrino spectra are not identical. If the DM and neutrino spectra are identical, the systematic uncertainty dominated regime persists indefinitely, resulting in a hard boundary on the DM cross section.

For semiconductor targets, DM-induced Migdal events and neutrino-induced Migdal events have similar spectra. However, this information is lost in our analysis since we have summed these events with \cevns and $\nu-e$ events, as they are indistinguishable within the recoil energy regime of interest. In our analysis $pp$ solar neutrinos are inaccessible to \cevns but can be accessed through $\nu-e$ scattering and the Migdal effect induced by neutrinos. Moreover, for semiconductor targets, given that Migdal-induced DM and neutrino events have identical spectra within our considered $n_e$ bins, there would theoretically be a hard floor. This is evident from the relatively flat behavior of the discovery limit over a large exposure range (see the orange and green lines in Fig.\,\ref{fig:DL_1D_all}). However, the inclusion of $\nu-e$ scattering introduces differences between the DM and neutrino spectra near the tail of the distribution, enabling separation of DM and neutrinos. This highlights the importance of including subdominant $\nu-e$ scattering in our analysis. As a result, theoretically, there would not be a hard floor in the cross section. With sufficiently high exposure, it would, in principle, be possible to discover small cross sections. Consequently, the term neutrino fog has been adopted to describe these neutrino background-dominated regions, as opposed to a hard boundary, which is referred to as the neutrino floor, following the formal proposal in\,\cite{OHare:2021utq}.

\begin{figure*}[t]
\begin{center}
	\subfloat[\label{sf:DL_2D_Si}]{\includegraphics[angle=0.0,width=0.45\textwidth]{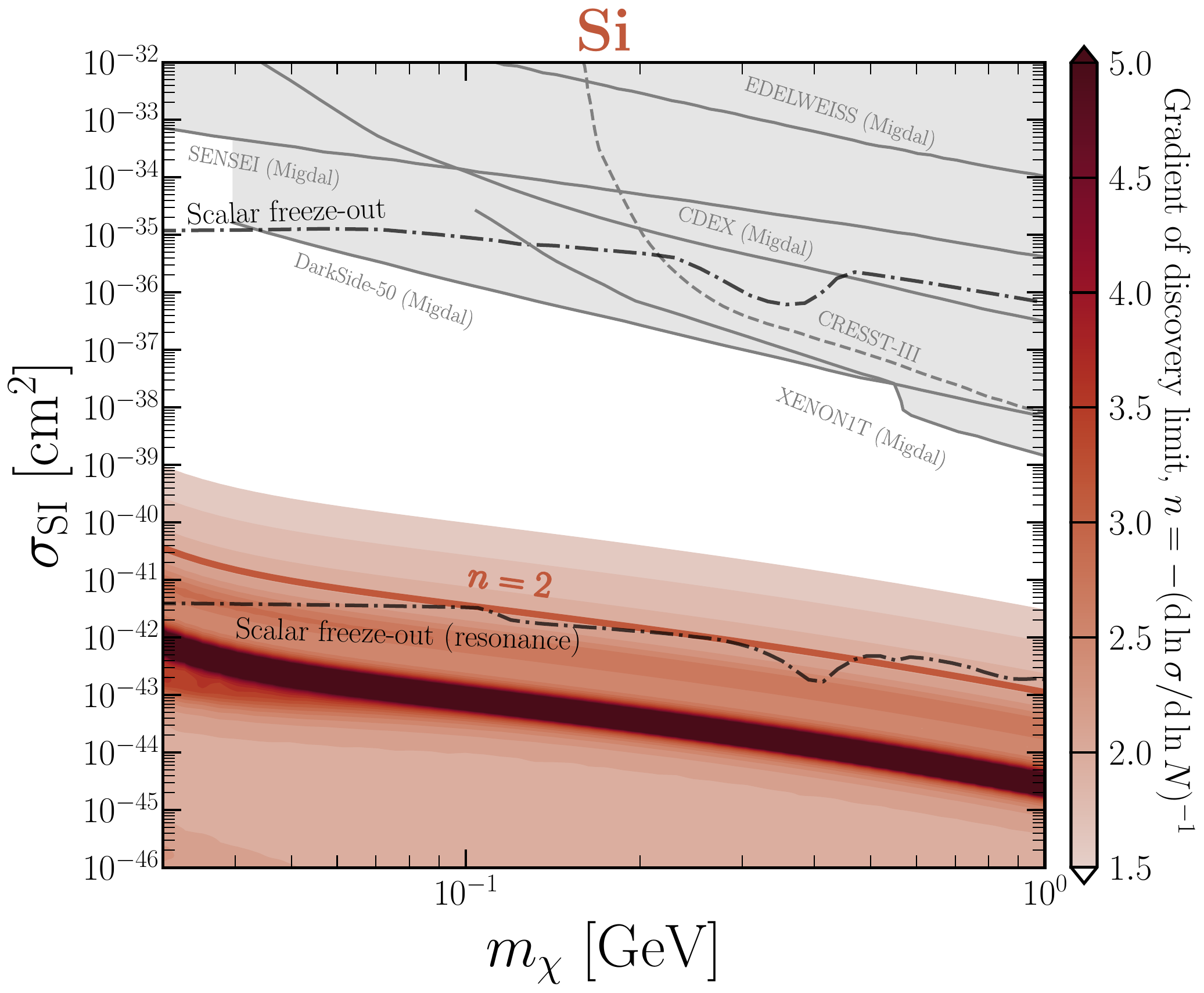}}~~
	\subfloat[\label{sf:DL_2D_Ge}]{\includegraphics[angle=0.0,width=0.45\textwidth]{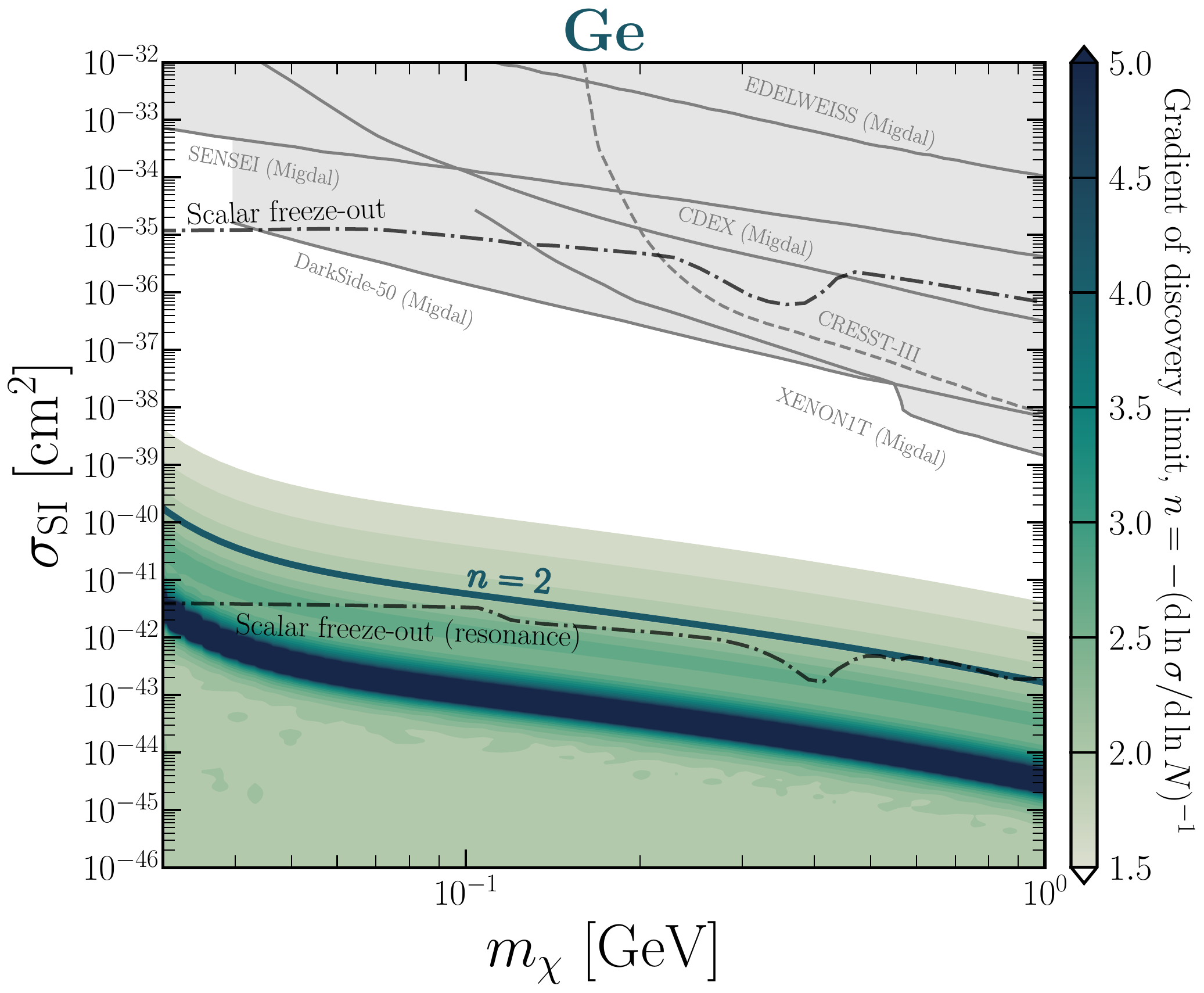}}~~\\	
	\caption{Same as Fig.\,\ref{fig:DL_2D_Xe_Ar} but for semiconductor (a) Si and (b) Ge targets.}
	\label{fig:DL_2D_Si_Ge}
\end{center}	
\end{figure*}

Following Ref.\, \cite{OHare:2021utq} we quantify fogginess though the gradient of discovery limit ($n$) with exposure using following equation:
\begin{equation}
n = - \dv{\ln \sigma_{\rm SI}}{\ln N}
\label{eq:gradient}
\end{equation}   
Notably, for very small exposures, i.e., the background-free search regime $n \lesssim 1$, in the statistical uncertainty-dominated regime $n = 2$, and in the systematic uncertainty-dominated regime $n \geq 2$. The values of $n$ are mapped in the DM mass and SI DM-nucleon cross section parameter space in Figs.\,\ref{sf:DL_2D_Xe}, \ref{sf:DL_2D_Ar}, \ref{sf:DL_2D_Si} and \ref{sf:DL_2D_Ge} for Xe, Ar, Si, and Ge, respectively. We also show the $n = 2$ boundary of the neutrino fog, where the DM search transits from the statistical uncertainty dominated regime to the systematic uncertainty dominated regime. Crossing this systematic uncertainty dominated regime would require an impractical amount of exposure, making this $n = 2$ boundary serve as a visual guide to the neutrino floor for Migdal searches. For semiconductors, we do not extend our analysis below DM masses of $\sim 30\,$MeV due to limitations in the applicability of ionization probabilities at lower masses.

In Figs.\,\ref{fig:DL_2D_Xe_Ar} and \ref{fig:DL_2D_Si_Ge}, we show the current Migdal bounds on the DM parameter space from XENON1T \cite{XENON:2019zpr}, DarkSide-50 \cite{DarkSide:2022dhx}, SENSEI \cite{SENSEI:2023zdf}, CDEX \cite{CDEX:2021cll} and EDELWEISS \cite{armengaud2022search} by the gray solid lines, along with the subdominant CRESST-II \cite{CRESST:2019jnq} bound from elastic nuclear scattering by the grey dashed line. Other sub-dominant bounds \cite{LUX:2018akb, PhysRevLett.116.071301, NEWS-G:2017pxg, Collar:2018ydf, Maity:2022exk, Super-Kamiokande:2022ncz, Mahdawi:2018euy, Buen-Abad:2021mvc, Rogers:2021byl, Das:2022srn, Das:2024jdz, COSINE-100:2021poy} are omitted for clarity. The regions between the black dot-dashed lines represent the scalar DM parameter space that satisfies the relic density. The upper dot-dashed line corresponds to typical annihilation setting the DM abundance, while the lower line corresponds to resonance annihilation, which results in a smaller scattering cross section \cite{Feng:2017drg}. These are for illustrative purposes; for other models, see Refs.\,\cite{Boehm:2003hm, Izaguirre:2015yja, Essig:2022dfa, Binder:2022pmf, Balan:2024cmq}. We note that for probing certain parts of the relic allowed parameter space, neutrinos would be an important background to tackle. 

Following our analysis, we find that both future Xe and Ar based experiments, which are expected to reach the exposures of $\mathcal{O}(10^5)$ kg-year \cite{DarkSide-20k:2024yfq, XLZD:2024nsu}, will enter the neutrino background-dominated regime in DM search using the Migdal effect. However, the same does not hold for silicon and germanium targets, primarily due to their significantly lower projected exposures of only $\mathcal{O}(10^2)$ kg-year \cite{Oscura:2022vmi}. In order to probe the neutrino background–dominated DM parameter space using the Migdal effect in future Xe and Ar based experiments, it would be highly beneficial to distinguish Migdal event topologies—particularly from \cevns events-potentially utilizing S1-S2 signatures. Another potential strategy involves exploring differences in the angular correlation between the recoiling nucleus and the emitted electron \cite{PhysRevA.15.162}, especially in directional experiments such as CYGNUS \cite{Vahsen:2020pzb}, for DM versus neutrino-induced events.

\section{Observing Migdal signal through neutrinos?}
\label{sec:nusignal}
In this section, we explore the feasibility of discovering neutrino-induced Migdal events in future DD experiments, neglecting DM. Our aim is to determine whether it is at all possible to observe Migdal events from neutrinos, which may warrant a more detailed study. To this end, we performed a rather aggressive analysis to estimate the exposure required to detect the neutrino generated Migdal event rate in the presence of other neutrino-only backgrounds, assuming a one-electron threshold. Unlike the previous case, we use the following test statistic for this analysis
\begin{equation}
q_0 = -2 \ln \frac{\mathcal{L}_{s+b}}{\mathcal{L}_{b}},
\label{eq:TS_discovery}
\end{equation}
where $\mathcal{L}_{s+b}$ is the likelihood for signal and background, and $\mathcal{L}_{b}$ is the likelihood for the background-only analysis. The signal consists of Migdal events produced by neutrinos, represented by the gray solid lines in Figs.\,\ref{fig:rate_xe_ar} and \ref{fig:rate_si_ge}. The background is the combination of \cevns and $\nu-e$ events from each of the neutrino sources. These are then input into a version of Eq.\,\eqref{eq:ll} to obtain each likelihood. We then numerically evaluate $q_0$ values for each exposure. The discovery significance for a given exposure is obtained by $\sqrt{q_0}$.

\begin{figure*}[t]
\begin{center}
	\subfloat[\label{sf:DS_Xe_Ar}]{\includegraphics[angle=0.0,width=0.45\textwidth]{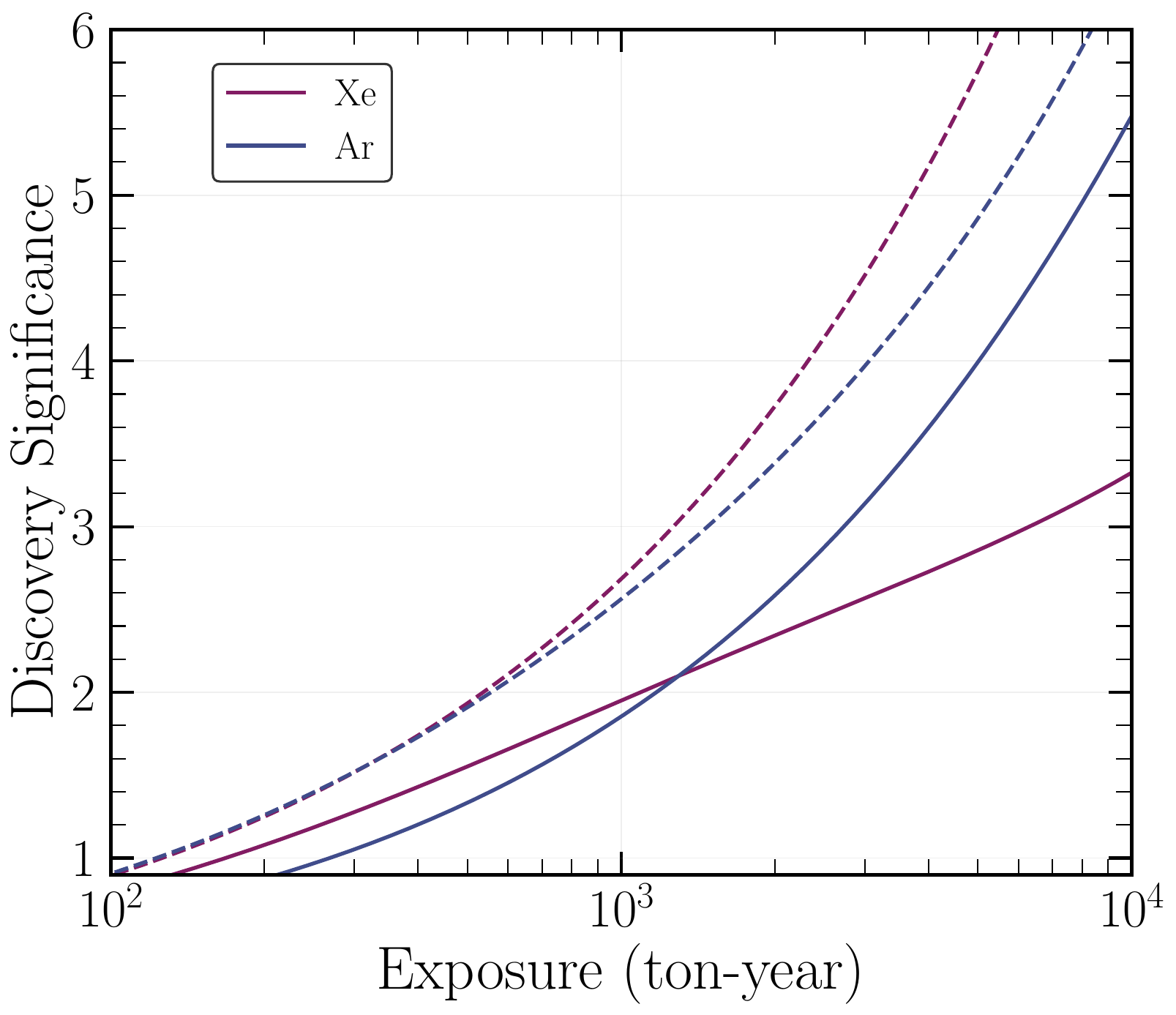}}~~
	\subfloat[\label{sf:DS_Si_Ge}]{\includegraphics[angle=0.0,width=0.435\textwidth]{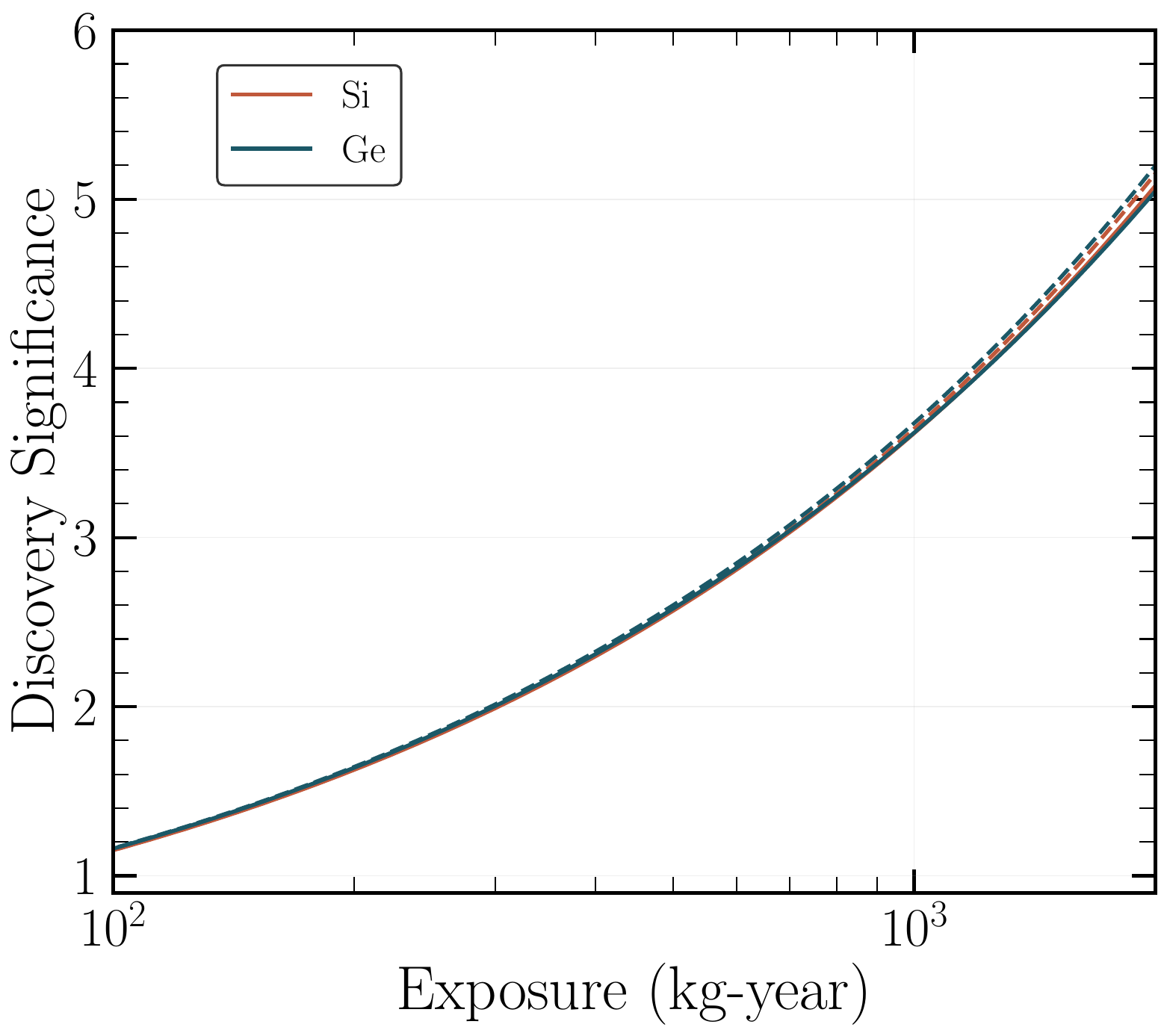}}~~\\	
	\caption{Neutrino-induced Migdal effect discovery significance as a function of exposure is shown. For the dashed lines considered neutrino flux uncertainties are reduced by a factor of $10$ compared to the solid lines. (a) The purple and blue lines correspond to Xe and Ar targets, respectively, with exposure measured in ton-years. (b) The orange and blue lines correspond to Si and Ge targets, respectively, with exposure measured in kg-years. }
	\label{fig:DS_Migdal}
\end{center}	
\end{figure*}

The discovery significance as a function of exposure for liquid noble and semiconductor targets is displayed in Figs.\,\ref{sf:DS_Xe_Ar} and \ref{sf:DS_Si_Ge}, respectively. In both plots, solid lines represent the discovery significance with neutrino flux uncertainties from \cite{OHare:2020lva}, while the dashed lines show the significance with neutrino flux uncertainties improved by a factor of 10. For similar significance, the required exposure for semiconductor detectors is much smaller than for Xe and Ar due to their larger Migdal rates. Since the neutrino Migdal rate in semiconductors is always greater than the background fluctuations, improving neutrino flux uncertainties does not enhance the discovery significance. As a result, the solid and dashed lines largely overlap in Fig.\,\ref{sf:DS_Si_Ge}. Unfortunately, for both cases, it will be difficult to achieve a discovery with future DD experiments. Note that for future Xe-based detectors (such as XLZD \cite{XLZD:2024nsu} or PandaX-xT  \cite{PANDA-X:2024dlo}) and Si-based detectors (like Oscura \cite{Oscura:2022vmi}), the exposure for a $10$-year running period is expected to be around $500$ ton-years for Xe detectors and $100$ kg-years for Si detectors.

\section{Conclusion}
\label{sec:conclusion}
Typical nuclear recoil DD experiments measure the recoil of nucleus caused by the elastic scattering of ambient nonrelativistic DM under the assumption that the electron clouds in the material instantly follow the nuclear recoil. In reality, this assumption is not entirely accurate; the electron cloud does not immediately follow the nuclear recoil. This nonadiabatic response of the electron cloud leads to ionization, a phenomenon known as the Migdal effect. It is particularly useful in the search for light DM, where the nuclear recoil remains below the detection threshold, but the accompanying ionization can produce observable signatures.    In fact, the Migdal effect currently sets the strongest bounds for the SI DM-nucleon scattering cross section in the DM mass range of approximately 0.01 --1 GeV, as established by various current DM DD experiments.

In this paper, we quantify the SI DM-nucleon parameter space that would be challenging to probe through the Migdal effect due to the presence of the neutrino background. Since the best bounds on the DM parameters of interest are derived from ionization-only analyses, we performed a similar analysis in our calculations. In such analyses, it is difficult to determine whether the ionization events arise from the quenching of nuclear recoil, electron recoil, or the Migdal effect. Therefore, in our neutrino background model, we included events from \cevns, $\nu-e$ scattering, and Migdal-induced neutrino events. We employed state-of-the-art profile likelihood techniques to map the neutrino fog for the SI DM-nucleon scattering parameter space. Additionally, we provide a visual guide to the neutrino floor—the SI DM-nucleon cross-section at which DD experiments move through from the neutrino statistical uncertainty dominated regime to the systematic uncertainty dominated regime. Based on our analysis, we conclude that future large-scale Xe and Ar target experiments are expected to be limited by the neutrino background, whereas this is not the case for the relatively smaller-scale Si and Ge target experiments. Our results show that it is essential to include all the aforementioned neutrino backgrounds to accurately quantify whether a hard floor exists or not. Furthermore, we find that neutrinos will be a significant background for probing parts of the relic density allowed parameter space in DM searches leveraging the Migdal effect.  

Finally, we have estimated the exposure required to observe neutrino-induced Migdal events in future DD experiments for all the considered targets. We quantify this in the presence of \cevns and $\nu-e$ scattering backgrounds. We find that it would be challenging to detect neutrino-induced Migdal events in future experiments. This is primarily due to the overwhelming \cevns backgrounds. This suggests that, to feasibly discover the Migdal effect using neutrinos, additional strategies are needed, including possibly distinguishing the Migdal event topology from other recoil mechanisms and potentially utilizing angular, timing  information in the production of ionized electrons.

\emph{Acknowledgments --} TNM thank Ranjan Laha, Jayden Newstead and Ciaran O'Hare for discussions and Rohan Pramanick for assistance with computations. The work of TNM is supported by the Australian Research Council through the ARC Centre of Excellence for Dark Matter Particle Physics.


\bibliographystyle{JHEP}
\bibliography{nu_migdal.bib}
\end{document}